\documentclass{elsart}
      \journal{Physics Letters {\bf B}}
      \usepackage{amssymb}
      \usepackage{graphicx}
      \usepackage{epsfig}  
      \begin{document}
      \begin{frontmatter}
	\vspace{0cm}
	\hspace{9.5cm}\mbox{CERN-PH-EP/2004-47} \\
	\hspace{8.8cm}\mbox{20 September 2004} \\
      \title{Measurement of the branching ratio of the decay $K_L
\rightarrow\pi^{\pm}e^{\mp}\nu$ and extraction of the CKM parameter $|V_{us}|$ }
      \date{}
\collab{NA48 Collaboration}
\author{A.~Lai},
\author{D.~Marras}
\address{Dipartimento di Fisica dell'Universit\`a e Sezione dell'INFN di Cagliari, \\ I-09100 Cagliari, Italy} 
\author{A.~Bevan},
\author{R.S.~Dosanjh\thanksref{threfRAL1}},
\author{T.J.~Gershon\thanksref{threfRAL2}},
\author{B.~Hay},
\author{G.E.~Kalmus},
\author{C.~Lazzeroni},
\author{D.J.~Munday},
\author{E.~Olaiya\thanksref{threfRAL}},
\author{M.A.~Parker},
\author{T.O.~White},
\author{S.A.~Wotton}
\address{Cavendish Laboratory, University of Cambridge, Cambridge, CB3~0HE, U.K.\thanksref{thref3}}
\thanks[thref3]{Funded by the U.K.\ Particle Physics and Astronomy Research Council}
\thanks[threfRAL1]{Present address: Ottawa-Carleton Institute for Physics, Carleton University, Ottawa, Ontario K1S 5B6, Canada}
\thanks[threfRAL2]{Present address: High Energy Accelerator Research Organization (KEK), Tsukuba, Japan}
\thanks[threfRAL]{Present address: Rutherford Appleton Laboratory, Chilton, Didcot, Oxon, OX11~0QX, U.K.}

\author{G.~Barr\thanksref{threfZX1}},
\author{G.~Bocquet},
\author{A.~Ceccucci},
\author{T.~Cuhadar-D\"onszelmann\thanksref{threfZX2}},
\author{D.~Cundy\thanksref{threfZX}},
\author{G.~D'Agostini},
\author{N.~Doble\thanksref{threfPisa}},
\author{V.~Falaleev},
\author{L.~Gatignon},
\author{A.~Gonidec},
\author{B.~Gorini},
\author{G.~Govi},
\author{P.~Grafstr\"om},
\author{W.~Kubischta},
\author{A.~Lacourt},
\author{A.~Norton},
\author{S.~Palestini},
\author{B.~Panzer-Steindel},
\author{H.~Taureg},
\author{M.~Velasco\thanksref{threfNW}},
\author{H.~Wahl\thanksref{threfHW}}
\address{CERN, CH-1211 Gen\`eve 23, Switzerland}
\thanks[threfZX1]{Present address: Department of Physics, University of Oxford, Denis Wilkinson Building, Keble Road, Oxford, UK, OX1 3RH}
\thanks[threfZX2]{Present address: University of British Columbia, Vancouver, BC, Canada, V6T 1Z1} 
\thanks[threfZX]{Present address: Istituto di Cosmogeofisica del CNR di Torino, I-10133~Torino, Italy}
\thanks[threfPisa]{Present address: Scuola Normale Superiore e Sezione dell'INFN di Pisa, I-56100~Pisa, Italy}
\thanks[threfNW]{Present address: Northwestern University, Department of Physics and Astronomy, Evanston, IL~60208, USA}
\thanks[threfHW]{Present address: Dipartimento di Fisica dell'Universit\`a e Sezione dell'INFN di Ferrara, I-44100~Ferrara, Italy}

\author{C.~Cheshkov\thanksref{threfCERN}},
\author{A.~Gaponenko},
\author{P.~Hristov\thanksref{threfCERN}},
\author{V.~Kekelidze},
\author{L.~Litov},
\author{D.~Madigojine},
\author{N.~Molokanova},
\author{Yu.~Potrebenikov},
\author{S.~Stoynev},
\author{G.~Tatishvili\thanksref{threfCM}},
\author{A.~Tkatchev},
\author{A.~Zinchenko}
\address{Joint Institute for Nuclear Research, Dubna, 141980, Russian Federation}  
\thanks[threfCERN]{Present address: PH Department, CERN, CH-1211 Geneva~23, Switzerland}
\thanks[threfCM]{Present address: Carnegie Mellon University, Pittsburgh, PA~15213, USA}
\author{I.~Knowles},
\author{V.~Martin\thanksref{threfNW}},
\author{R.~Sacco\thanksref{threfSacco}},
\author{A.~Walker}
\address{Department of Physics and Astronomy, University of Edinburgh, JCMB King's Buildings, Mayfield Road, Edinburgh, EH9~3JZ, U.K.} 
\thanks[threfSacco]{Present address: Department of Physics, Queen Mary, University of London, Mile End Road, London, E1 4NS}
\newpage
\author{M.~Contalbrigo},
\author{P.~Dalpiaz},
\author{J.~Duclos},
\author{P.L.~Frabetti\thanksref{threfFrabetti}},
\author{A.~Gianoli},
\author{M.~Martini},
\author{F.~Petrucci},
\author{M.~Savri\'e}
\address{Dipartimento di Fisica dell'Universit\`a e Sezione dell'INFN di Ferrara, I-44100 Ferrara, Italy}
\thanks[threfFrabetti]{Present address: Joint Institute for Nuclear Research, Dubna, 141980, Russian Federation}
\author{A.~Bizzeti\thanksref{threfXX}},
\author{M.~Calvetti},
\author{G.~Collazuol\thanksref{threfPisa}},
\author{G.~Graziani\thanksref{threfGG}},
\author{E.~Iacopini},
\author{M.~Lenti},
\author{F.~Martelli\thanksref{thref7}},
\author{M.~Veltri\thanksref{thref7}}
\address{Dipartimento di Fisica dell'Universit\`a e Sezione dell'INFN di Firenze, I-50125~Firenze, Italy}
\thanks[threfXX]{Dipartimento di Fisica dell'Universit\`a di Modena e Reggio Emilia, I-41100~Modena, Italy}
\thanks[threfGG]{Present address: DSM/DAPNIA - CEA Saclay, F-91191 Gif-sur-Yvette, France}
\thanks[thref7]{Istituto di Fisica dell'Universit\`a di Urbino, I-61029~Urbino, Italy}
\author{H.G.~Becker},
\author{K.~Eppard},
\author{M.~Eppard\thanksref{threfCERN}},
\author{H.~Fox\thanksref{threfNW}},
\author{A.~Kalter},
\author{K.~Kleinknecht},
\author{U.~Koch},
\author{L.~K\"opke},
\author{P.~Lopes da Silva}, 
\author{P.~Marouelli},
\author{I.~Pellmann\thanksref{threfDESY}},
\author{A.~Peters\thanksref{threfCERN}},
\author{B.~Renk},
\author{S.A.~Schmidt},
\author{V.~Sch\"onharting},
\author{Y.~Schu\'e},
\author{R.~Wanke},
\author{A.~Winhart},
\author{M.~Wittgen\thanksref{threfSLAC}}
\address{Institut f\"ur Physik, Universit\"at Mainz, D-55099~Mainz, Germany\thanksref{thref6}}
\thanks[thref6]{Funded by the German Federal Minister for Research and Technology (BMBF) under contract 7MZ18P(4)-TP2}
\thanks[threfDESY]{Present address: DESY Hamburg, D-22607~Hamburg, Germany}
\corauth[cor]{Corresponding author.\\{\em Email address:} burkhard.renk@uni-mainz.de}
\thanks[threfSLAC]{Present address: SLAC, Stanford, CA~94025, USA}
\author{J.C.~Chollet},
\author{L.~Fayard},
\author{L.~Iconomidou-Fayard},
\author{J.~Ocariz},
\author{G.~Unal},
\author{I.~Wingerter-Seez}
\address{Laboratoire de l'Acc\'el\'erateur Lin\'eaire, IN2P3-CNRS,Universit\'e de Paris-Sud, 91898 Orsay, France\thanksref{threfOrsay}}
\thanks[threfOrsay]{Funded by Institut National de Physique des Particules et de Physique Nucl\'eaire (IN2P3), France}
\author{G.~Anzivino},
\author{P.~Cenci},
\author{E.~Imbergamo},
\author{P.~Lubrano},
\author{A.~Mestvirishvili},
\author{A.~Nappi},
\author{M.~Pepe},
\author{M.~Piccini}
\address{Dipartimento di Fisica dell'Universit\`a e Sezione dell'INFN di Perugia, \\ I-06100 Perugia, Italy}
\author{R.~Casali},
\author{C.~Cerri},
\author{M.~Cirilli\thanksref{threfCERN}},
\author{F.~Costantini},
\author{R.~Fantechi},
\author{L.~Fiorini},
\author{S.~Giudici},
\author{G.~Lamanna},
\author{I.~Mannelli},
\author{G.~Pierazzini},
\author{M.~Sozzi}
\address{Dipartimento di Fisica, Scuola Normale Superiore e Sezione dell'INFN di Pisa, \\ I-56100~Pisa, Italy} 
\author{J.B.~Cheze},
\author{J.~Cogan},
\author{M.~De Beer},
\author{P.~Debu},
\author{A.~Formica},
\author{R.~Granier de Cassagnac\thanksref{threfEcolePoly}},
\author{E.~Mazzucato},
\author{B.~Peyaud},
\author{R.~Turlay},
\author{B.~Vallage}
\address{DSM/DAPNIA - CEA Saclay, F-91191 Gif-sur-Yvette, France} 
\thanks[threfEcolePoly]{Present address: Laboratoire Leprince-Ringuet,
\'Ecole polytechnique (IN2P3, Palaiseau, 91128 France}

%
%
%
\author{M.~Holder},
\author{A.~Maier\thanksref{threfCERN}},
\author{M.~Ziolkowski}
\address{Fachbereich Physik, Universit\"at Siegen, D-57068 Siegen, Germany\thanksref{thref8}}
\thanks[thref8]{Funded by the German Federal Minister for Research and Technology (BMBF) under contract 056SI74}
\author{R.~Arcidiacono},
\author{C.~Biino},
\author{N.~Cartiglia},
\author{R.~Guida}, 
\author{F.~Marchetto}, 
\author{E.~Menichetti},
\author{N.~Pastrone}
\address{Dipartimento di Fisica Sperimentale dell'Universit\`a e Sezione dell'INFN di Torino, I-10125~Torino, Italy} 
\author{J.~Nassalski},
\author{E.~Rondio},
\author{M.~Szleper\thanksref{threfNW}},
\author{W.~Wislicki},
\author{S.~Wronka}
\address{Soltan Institute for Nuclear Studies, Laboratory for High Energy Physics, PL-00-681~Warsaw, Poland\thanksref{thref9}}
\thanks[thref9]{Supported by the KBN under contract SPUB-M/CERN/P03/DZ210/2000 and using computing resources of the
Interdisciplinary Center for Mathematical and Computational Modelling of the University of Warsaw.}
\author{H.~Dibon},
\author{G.~Fischer},
\author{M.~Jeitler},
\author{M.~Markytan},
\author{I.~Mikulec},
\author{G.~Neuhofer},
\author{M.~Pernicka},
\author{A.~Taurok},
\author{L.~Widhalm}
\address{\"Osterreichische Akademie der Wissenschaften, Institut f\"ur Hochenergiephysik, A-1050~Wien, Austria\thanksref{thref10}}
\thanks[thref10]{Funded by the Federal Ministry of Science and Transportation under the contract GZ~616.360/2-IV GZ 616.363/2-VIII, 
and by the Austrian Science Foundation under contract P08929-PHY.}

     \begin{abstract}
We present a new measurement of the branching ratio $R$ of the decay $K_L
\rightarrow\pi^{\pm}e^{\mp}\nu$, denoted as $K_{e3}$, relative to all charged 
$K_L$ decays with two tracks, based on data taken with the NA48 detector at the CERN
SPS. We measure $R = 0.4978 \pm 0.0035$. From this we derive the
$K_{e3}$ branching fraction and the weak coupling parameter $|V_{us}|$
in the CKM matrix. We obtain $|V_{us}|f_{+}(0) = 0.2146 \pm 0.0016 $,
where $f_{+}(0)$ is the vector form factor in the $K_{e3}$ decay.
\end{abstract}

     \end{frontmatter}
     \setcounter{footnote}{0}


\newcommand{\vet}{\protect\varepsilon_{\theta\theta}}
\newcommand{\vez}{\protect\varepsilon_{zz}} \pagestyle{headings}


\renewcommand{\topfraction}{1.}  \renewcommand{\bottomfraction}{1.}
\renewcommand{\textfraction}{0.}  \renewcommand{\floatpagefraction}{1.}

\newcommand{\egKK}{e.g.}  \newcommand{\eKK}{\mbox{$e$}}
\newcommand{\mrm}{\mathrm} \newcommand{\bfS}{{\bf{S}}}
\newcommand{\CP}{\mbox{$\mathrm{CP}$}}
\newcommand{\eps}{\mbox{$\epsilon$}}
\newcommand{\epsp}{\mbox{$\epsilon'$}}
\newcommand{\epsbar}{\mbox{$\bar{\epsilon}$}}
\newcommand{\eoe}{\mbox{$\re(\epsp/\eps)$}}
\newcommand{\ese}{\mbox{$\epsp/\eps$}}
\newcommand{\kz}{\mbox{$\mathrm{K^0}$}}
\newcommand{\kl}{\mbox{$\mathrm{K_L}$}}
\newcommand{\kzbar}{\mbox{$\mathrm{\overline{K^0}}$}}
\newcommand{\kethree}{\mbox{$\mathrm{K_{e3}}$}}
\newcommand{\kmut}{\mbox{$\mathrm{K_{\mu 3}}$}}
\newcommand{\kpit}{\mbox{$\mathrm K_{\pi3}$}}
\newcommand{\ks}{\mbox{$\mathrm{K_S}$}}
\newcommand{\ksl}{\mbox{$\mathrm{K_{S,L}}$}}
\newcommand{\pim}{\mbox{$\mrm\pi^{-}$}}
\newcommand{\pip}{\mbox{$\mrm\pi^{+}$}}
\newcommand{\piz}{\mbox{$\mrm\pi^{0}$}}
\newcommand{\rell}{\mbox{$\mathrm R_{\mathrm ell}$}}
\newcommand{\kstocpipi}{\mbox{$\ks \rightarrow\pi^+\pi^-$}}
\newcommand{\kltocpipi}{\mbox{$\kl \rightarrow\pi^+\pi^-$}}
\newcommand{\ktocpipi}{\mbox{$\mathrm{K}\rightarrow\pi^+\pi^-$}}
\newcommand{\kstonpipi}{\mbox{$\ks \rightarrow\pi^0\pi^0$}}
\newcommand{\kltonpipi}{\mbox{$\kl \rightarrow\pi^0\pi^0$}}
\newcommand{\ktonpipi}{\mbox{$\mathrm{K}\rightarrow\pi^0\pi^0$}}
\newcommand{\im}{\mbox{$\mathcal{I}m$}}
\newcommand{\re}{\mbox{$\mathcal{R}e$}}
\newcommand{\isoxi}[1]{\mbox{$\frac{\im A_{#1}}{\re A_{#1}}$}}
\newcommand{\rea}[1]{\mbox{$\re A_{#1}$}}
\newcommand{\ima}[1]{\mbox{$\im A_{#1}$}}
\newcommand{\minus}[1]{\mbox{$-#1$}} \newcommand{\ketKK}[2]{$|#1 \  #2
\rangle$} \newcommand{\etapm}{\mbox{$\eta_{+-}$}}
\newcommand{\etazz}{\mbox{$\eta_{00}$}}
\newcommand{\XPT}{\mbox{$\mrm{\chi PT}$} }
\newcommand{\als}{\mbox{$\alpha_{LS}$}}
\newcommand{\asl}{\mbox{$\alpha_{SL}$}}
\newcommand{\alspm}{\mbox{$\alpha_{LS}^{+-}$}}
\newcommand{\aslpm}{\mbox{$\alpha_{SL}^{+-}$}}
\newcommand{\alszz}{\mbox{$\alpha_{LS}^{00}$}}
\newcommand{\aslzz}{\mbox{$\alpha_{SL}^{00}$}}
\newcommand{\dals}{\mbox{$\Delta\als$}}
\newcommand{\dasl}{\mbox{$\Delta\asl$}}
\newcommand{\ptprime}{\mbox{$p_{\mathrm{t}}'$}}
\newcommand{\ptprimesq}{\mbox{$p_{\mathrm{t}}'^2$}}
\newcommand{\tten}[1]{\mbox{$\times 10^{#1}$}}

\newcommand{\laba}[1]{\label{sec:#1}}
\newcommand{\labc}[1]{\label{sec:#1}}
\newcommand{\labe}[1]{\label{equ:#1}}
\newcommand{\labs}[1]{\label{sec:#1}}
\newcommand{\labf}[1]{\label{fig:#1}}
\newcommand{\labt}[1]{\label{tab:#1}}
\newcommand{\refa}[1]{\ref{sec:#1}}
\newcommand{\refc}[1]{\ref{sec:#1}}
\newcommand{\refe}[1]{\ref{equ:#1}}
\newcommand{\refs}[1]{\ref{sec:#1}}
\newcommand{\reff}[1]{\ref{fig:#1}}
\newcommand{\reft}[1]{\ref{tab:#1}}
\newcommand{\eq}[1]{(\refe{#1})}
\newcommand{\Eq}[1]{Gleichung~\refe{#1}}
\newcommand{\Eqs}[1]{Eqs.~(\refe{#1})}
\newcommand{\Eqss}[2]{Eqs.~(\refe{#1}) and (\refe{#2})}
\newcommand{\Eqsss}[3]{Eqs.~(\refe{#1}), (\refe{#2}), and (\refe{#3})}
\newcommand{\Fig}[1]{Fig.~\reff{#1}}
\newcommand{\fig}[1]{fig.~\reff{#1}}
\newcommand{\Figs}[1]{Figs.~\reff{#1}}
\newcommand{\Figsss}[3]{Figs.~\reff{#1}, \reff{#2}, and \reff{#3}}
\newcommand{\Section}[1]{Section~\refs{#1}}
\newcommand{\Anh}[1]{Anhang~\refa{#1}}
\newcommand{\Chap}[1]{Kapitel~\refc{#1}}
\newcommand{\Sec}[1]{Abschnitt~\refs{#1}}
\newcommand{\Secs}[1]{Sects.~\refs{#1}}
\newcommand{\Secss}[2]{Abschnitte~\refs{#1} und \refs{#2}}
\newcommand{\Secsss}[3]{Sects.~\refs{#1}, \refs{#2}, and \refs{#3}}
\newcommand{\tab}[1]{table~\reft{#1}}
\newcommand{\Tab}[1]{Table~\reft{#1}}
\newcommand{\Tables}[1]{Tabellen~\reft{#1}}
\newcommand{\Tabless}[2]{Tabellen~\reft{#1} und \reft{#2}}
\newcommand{\Tablesss}[3]{Tabellen~\reft{#1}, \reft{#2}, and \reft{#3}}

\newcommand{\keywords}{}
\newcommand{\PACS}{61.43.--j, 71.30.+h, 73.40.Hm}

%
\section{Introduction}

The unitary condition for the first row of the CKM quark mixing matrix
is at present fulfilled only at the 10\% C. L.  \cite{PDG}. This has
renewed interest in the measurement of the coupling constant $V_{us}$
for strangeness-changing weak transitions. The most precise
information on $V_{us}$ comes from the decay $K_L \rightarrow \pi^{\pm} e^{\mp}
\nu$, which is a vector transition, and therefore is protected from
SU(3) breaking effects by the Ademollo-Gatto theorem \cite{AdeGato}. We present here
a new measurement with improved experimental precision.

%
\section{Apparatus}

The experiment was performed using the NA48 detector in a beam of
long-lived neutral kaons produced at the 450\,GeV proton synchrotron
SPS at CERN. The neutral $K_L$ beam was derived at an angle of 2.4\,mrad 
from an extracted proton beam hitting a beryllium target. The decay region 
starts at the exit face of the last of three collimators 126\,m downstream 
of the target. The experiment was originally designed and used for the
precision measurement of direct CP violation in kaon decays
\cite{Alai}. We report here on a study of semileptonic decays, 
for which data were taken in a pure $K_L$ beam in september 1999. The main 
elements of the detector relevant for this
exposure are the following:

The magnetic spectrometer consists of four drift
chambers (DCH), each with 8 planes of sense wires oriented along four
projections, each one rotated by 45 degrees with respect to the
previous one. The spatial resolution achieved per projection is
100\,$\mu m$, and the time resolution for an event is 0.7\,ns. The
volume between the chambers is filled with helium near atmospheric
pressure. The spectrometer magnet is a dipole with a field  integral
of 0.883\,Tm and is placed after the first two chambers. The distance
between the first and the last chamber is 21.8 meters. The spectrometer 
is designed to measure the momenta of the
charged particles with high precision - the momentum resolution is
given by
\begin{equation}
\sigma(p)/p = 0.48\,\% \oplus 0.009 \cdot p\,\%
\end{equation}
where $p$ is in GeV/c. 

The hodoscope is placed downstream from the last drift chamber. It
consists of two planes of scintillators segmented in horizontal and
vertical strips and arranged in four quadrants. The signals are used
for a fast coincidence of two charged particles in the trigger. The
time resolution from the hodoscope is $\approx 200$\,ps per track.

The electromagnetic calorimeter (Lkr) is a quasi-homogeneous
calorimeter based on liquid krypton, with tower readout.  The 13212
readout cells have cross sections of $ \approx 2 \times 2$\,cm$^{2}$. The
electrodes extend from the front to the back of the detector in a
small angle accordion geometry. The Lkr calorimeter measures the
$e^{\pm}$ and $\gamma$ energies by summing the ionization from their
electromagnetic showers. The energy resolution is:
\begin{equation}
\sigma(E)/E = 3.2\,\%/\sqrt{E}\oplus 9.0\,\%/E \oplus 0.42\,\%
\end{equation}
where $E$ is in GeV.

Charged decays were
triggered with a two-level trigger system. The trigger requirements
were two charged particles  in the scintillator hodoscope or in the
drift chambers coming from a vertex in the decay region.

A more detailed description of the NA48 setup can be found elsewhere
\cite{Alai}.

%
\section{Data analysis}

\subsection{Analysis strategy and events selection}

The basic quantity measured in this experiment is the ratio $R$ of decay
rates of $K_{e3}$ decays relative to all decays with two charged
particles in the final state, mainly $\pi e \nu$, $\pi \mu \nu$
(called $K_{\mu 3}$), $\pi^+\pi^-\pi^0$ (called $K_{3\pi}$),
$\pi^+\pi^-$ (called $K_{2\pi}$) and $3\pi^0$ with Dalitz decay of one $\pi^0$, 
denoted as $\pi^0\pi^0 ee\gamma$ or $\pi^0\pi^0\pi^0_D$. Since the
neutral decay modes to $3 \pi^0, 2 \pi^0$ and $\gamma \gamma$ have been
measured, and the correction for events with four tracks $B(4T)$ is
small, the sum of branching ratios of all $K_L$ decay modes with two
charged tracks $B(2T)$ is experimentally known \cite{PDG}

%
\begin{eqnarray}
B(2T) & \,=\, & 1 - \frac{\Gamma(K_L \rightarrow all~neutral)}{\Gamma(K_L
\rightarrow all)} - B(4T) \nonumber \\
 & \,=\, & 1-B(3 \pi^0)-B(2 \pi^0)-B(\gamma \gamma) + B(\pi^0\pi^0\pi^0_D) -
B(4T) \nonumber \\
 & \,=\, & 1.0048 - B(3 \pi^0)\,.
\end{eqnarray}
Since contributions from $K_S$ meson or $\Lambda$ hyperon decays and
from rare $K_L$ decays are negligible, less than $ 2 \times 10^{-5}$,
we obtain the $K_{e3}$ branching ratio using $B(2T)$:
\begin{equation}
B(e3)=\frac{\Gamma(K_{e3})}{\Gamma(K_L \rightarrow all)
}=\frac{\Gamma(K_{e3})}{\Gamma(K_L \rightarrow all~\mbox{2-}track)}
\times B(2T)\,.
\end{equation}

In this experiment, we therefore measure the ratio of $K_{e3}$ events
to all 2-track events $N_{2T}$ divided by their acceptances $a_e$
or $a_{2T}$ respectively:
\begin{equation}
R = \frac{N_e/a_e}{N_{2T}/a_{2T}}\,. \label{eq:R}
\end{equation}

Both numbers, $N_e$ and $N_{2T}$ are extracted from the same sample of
about 80 million recorded 2-track events. These were reconstructed
and subjected to offline filtering.

In the basic selection, two tracks were required with opposite charge
and a distance of closest approach below 3\,cm. The vertex fiducial
volume was defined to be between 8\,m and 33\,m from the final
collimator, and within 3\,cm of the beam-line. Events with high hit multiplicity were
rejected by requiring that no overflow condition occurred in the drift chambers. 
An overflow is generated if more than seven hits in a plane were recorded within
100\,ns. These cuts were passed by 48.795 million events.

Events were rejected if the time difference between the tracks
exceeded 6\,ns. Both tracks were required to be inside the detector
acceptance and within the momentum interval 10\,GeV/c to 120\,GeV/c. 
In order to allow a clear separation of pion and electron showers, we 
required the distance between the entry points of the two tracks at
the front face of the electromagnetic calorimeter to be larger than 25\,cm. 

The last selection criterion was applied to a measure of the kaon momentum, to avoid 
the region below 50\,GeV/c which is simulated inadequately. We used the sum of the 
moduli of the two momenta $P = P_1 + P_2$ for all decays. As a result 12.592 million events with 
$P > 60$\,GeV/c remained. For the denominator, no identification of individual 
decay modes was applied, and the average acceptance $a_{2T}$ applies to the 
requirements listed up to this point.

For the numerator $N_e$, the $K_{e3}$ signal was selected by a single
additional criterion that at least one track should be consistent with
an electron. This was done by requiring that the ratio $E/p$ exceed
0.93, where $E$ is the measured energy in the calorimeter and $p$ is the
measured momentum in the magnetic spectrometer. 6.759 million events were
accepted. The quantity $E/p$ is shown in \Fig{eop_ke3} for all tracks
of these $K_{e3}$ events. 
\begin{figure}[h]
  \begin{center}
    \includegraphics[width=11.cm]{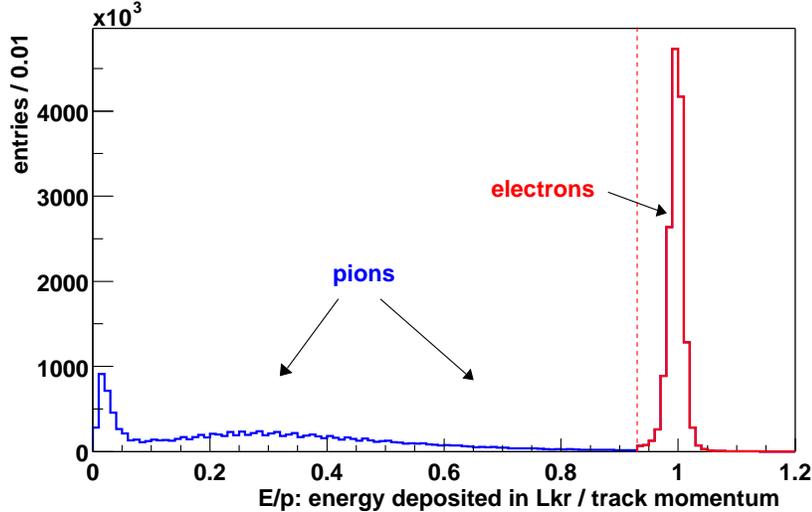}
    \parbox{11cm}{\caption[]
      {\labf{eop_ke3}The ratio of calorimetric energy $E$ over the
      momentum $p$ for the tracks of all selected $K_{e3}$ events.}}
  \end{center}
\end{figure}

\subsection{Corrections for electron identification}
The number of $K_{e3}$ events was corrected for the inefficiency of 
the electron identification (electrons with $E/p <
0.93$) and background coming from $K_{\mu 3}$ and $K_{3\pi}$ decays
(pions with $E/p > 0.93$). Both inefficiency and background were 
measured from the data.

For the background determination a sample of events was selected
having one track with $E/p > 1.0$, clearly classifying it as an
electron. The background probability for pions $W(\pi
\rightarrow e)$ was then determined from the $E/p$ spectrum of the
other (i.e. pion) track (see \Fig{eop_pion}) to be 
$$
W(\pi \rightarrow e) = (0.576 \pm 0.005 (stat.))\,\%. 
$$

As a cross check the probability was also derived from the $E/p$ spectrum of
$K_{3\pi}$ events, giving a consistent result within
errors. Background from the decay $K_L \rightarrow \pi^0\pi^0\pi^0_D$
was completely removed by the cut on $P$.

\begin{figure}[h]
  \begin{center}
    \includegraphics[width=11.cm]{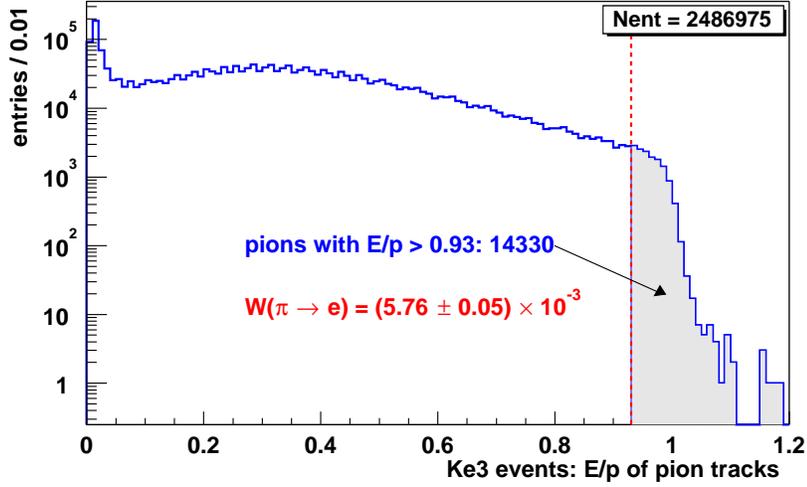}
    \parbox{11.cm}{\caption[]
      {\labf{eop_pion}Quantity $E/p$ for pion tracks. The sample was
      selected by the requirement $E/p > 1.0$ for the other (i.e. electron) track.}}
  \end{center}
\end{figure}
\begin{figure}[h]
  \begin{center}
    \includegraphics[width=11.cm]{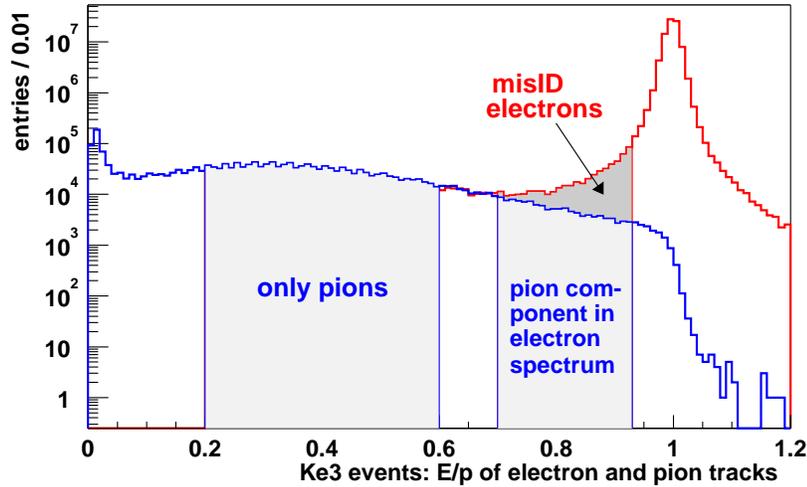}
    \parbox{11.cm}{\caption[]
      {\labf{eop_elec_pion}Quantity $E/p$ for electron and for pion
      tracks. The electron spectrum is scaled for better
      illustration. The dark shaded area represents electrons with
      $E/p < 0.93$.}}
  \end{center}
\end{figure}

The electron ID inefficiency $W(e \rightarrow \pi)$ was determined in a 
similar way (see \Fig{eop_elec_pion}) by requiring one track with
$E/p < 0.7$, classifying it as a pion. The $E/p$ distribution for the other
track then consists mainly of electrons, with a small contribution
from pions, especially below 0.7\,. We subtracted this pion component by
using the previously determined pion distribution, normalized in the
range $0.2 < E/p < 0.6$. From this we then obtained the
probability for losing an electron by the condition $E/p > 0.93$:
$$
W(e \rightarrow \pi) = (0.487 \pm 0.004 (stat.))\,\%.
$$

\subsection{Monte Carlo Simulation}

To reproduce the detector response, a GEANT-based simulation of the
NA48 apparatus was employed for the five decay modes $\pi e \nu$, $\pi
\mu \nu$,
$\pi^+\pi^-\pi^0$, $\pi^+\pi^-$ and $\pi^0\pi^0\pi^0_D$. Radiative
corrections were included for the $K_{e3}$ mode. We used the PHOTOS
program package \cite{Photos} to simulate bremsstrahlung, and
added the calculations from \cite{Gins} on virtual photons and
electrons. Some comparisons between data and MC for identified
$K_{e3}$ events are shown in \Fig{zvertex} (z-vertex) and
\Fig{track_dch1} (x- and y-coordinates of the tracks in the first drift chamber).
\begin{figure}[h]
  \begin{center}
    \includegraphics[width=13cm]{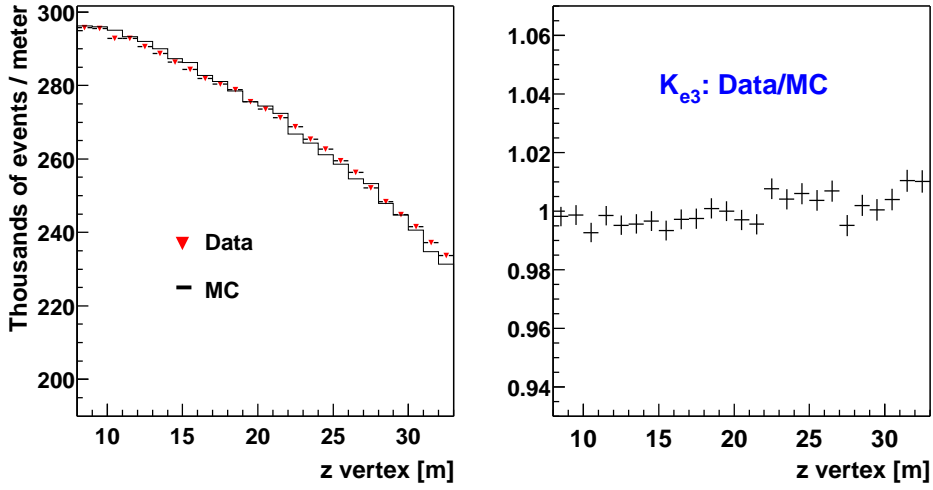}
    \parbox{12.5cm}{\caption[]
      {\labf{zvertex}Longitudinal vertex distribution for $K_{e3}$
      events: data and MC (left) and ratio of data over Monte Carlo
      simulation (right).}}
  \end{center}
\end{figure}

\begin{figure}[h]
  \begin{center}
    \includegraphics[width=12.5cm]{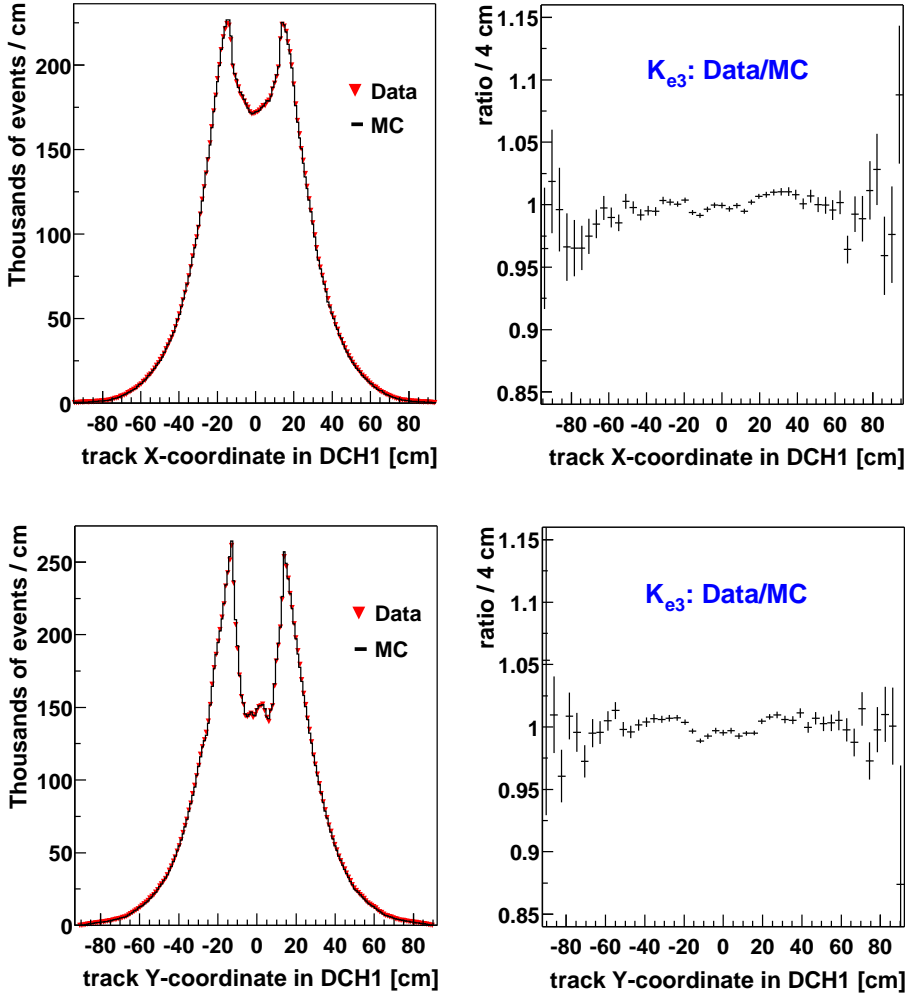}
    \parbox{12cm}{\caption[]
      {\labf{track_dch1}Transverse positions (horizontal x and
      vertical y) of tracks in the first drift chamber from $K_{e3}$
      events: data and MC (left) and ratio of data over Monte Carlo
      simulation (right).}}
  \end{center}
\end{figure}

We obtain the individual acceptances $a_i$ as shown in Table
\ref{tab:DetectorAcceptancesForTheChargedDecayModes}.

\begin{table}[ht]
    \begin{center}
	    \begin{tabular}[t]{|l|l|}\hline
    decay mode & acceptance\\	\hline 
    $K_{e3}$ & 0.2599 \\ 
    $K_{\mu3}$ & 0.2849 \\ 
    $K_{3\pi}$ & 0.0975 \\
    $K_{2\pi}$ & 0.5229 \\
    $K_{3\pi^0_D}$ & 0.0001 \\	 \hline
	    \end{tabular}
    \caption{Detector acceptances for the charged decay modes.}
    \label{tab:DetectorAcceptancesForTheChargedDecayModes}
    \end{center}
\end{table}

The average 2-track acceptance $a_{2T}$ was obtained from a weighted mean of
the individual acceptances $a_i$ which depends only on ratios of decay rates
measured in other experiments:
\begin{eqnarray}
a_{2T} & \,=\, & \frac{B_e a_e + B_\mu a_\mu + B_{3\pi} a_{3\pi} + B_{2\pi} a_{2\pi} +
B_D a_D}{B_e + B_\mu+ B_{3\pi} + B_{2\pi}+ B_D} \nonumber\\
 & \,=\, & \frac{a_e ( 1 + \frac{B_\mu}{B_e} \frac{a_\mu}{a_e} +
\frac{B_{3\pi}}{B_e} \frac{a_{3\pi}}{a_e} +  \frac{B_{2\pi}}{B_e}
\frac{a_{2\pi}}{a_e} +  \frac{B_D}{B_e} \frac{a_D}{a_e})}{( 1 +
\frac{B_\mu}{B_e} +  \frac{B_{3\pi}}{B_e} +  \frac{B_{2\pi}}{B_e} +
\frac{B_D}{B_e})} . \label{eq:Acc2t}
\end{eqnarray}

Here $B_i$ are the branching ratios for the decay channels
$(i$\,=\,$e: K_{e3}$; $i$\,=\,$\mu: K_{\mu3}$; $i$\,=\,${3\pi}:
\pi^+\pi^-\pi^0$; $i$\,=\,$2\pi: \pi^+\pi^-$; $i$\,=\,$D: \pi^0\pi^0\pi^0_D)$. The acceptance for
channel $i$ is $a_i$. For the branching ratios we used a weighted
average of the 2004 PDG values \cite{PDG} and the new KTeV measurement
\cite{KTEV}. The uncertainty was enlarged according to PDG rules for
averaging inconsistent data.
\begin{eqnarray}
B_\mu/B_e & \,=\, & 0.666 \pm 0.011 \quad  \label{eq:ExR1} \\
B_{3\pi}/B_e & \,=\, & 0.309 \pm 0.004 \quad \label{eq:ExR2} \\
B_{2\pi}/B_e & \,=\, & ( 4.90 \pm 0.14 ) \times 10^{-3} \label{eq:ExR10} \\
B_D/B_e & \,=\, & (1.96 \pm 0.05 ) \times 10^{-2} \label{eq:ExR11}
\end{eqnarray}
Varying the constraints given by Eq. (\ref{eq:ExR1}) to
(\ref{eq:ExR11}) within their errors we get a relative variation
of the acceptance of $0.16\,\%$, and $a_{2T} = 0.2412 \pm 0.0004$.
\subsection{Systematic uncertainties}

Given the large number of events, the uncertainties of this
measurement are purely of systematic nature. Simulation shows that most of these
systematics will induce a dependence of the result on the lower cut on
the sum of the two moduli of the two momenta $P = P_1 + P_2$. Since
the three decay modes have different neutral energy, which is either
invisible as a neutrino or not used in this analysis, events with a
given value of $P$ originate from different average kaon energies, so a possible
imperfection of the kaon energy spectrum (which is fairly well known
for energies above 50 GeV) will induce a dependence of the result on $P$.

\begin{figure}[b]
  \begin{center}
    \includegraphics[width=12cm]{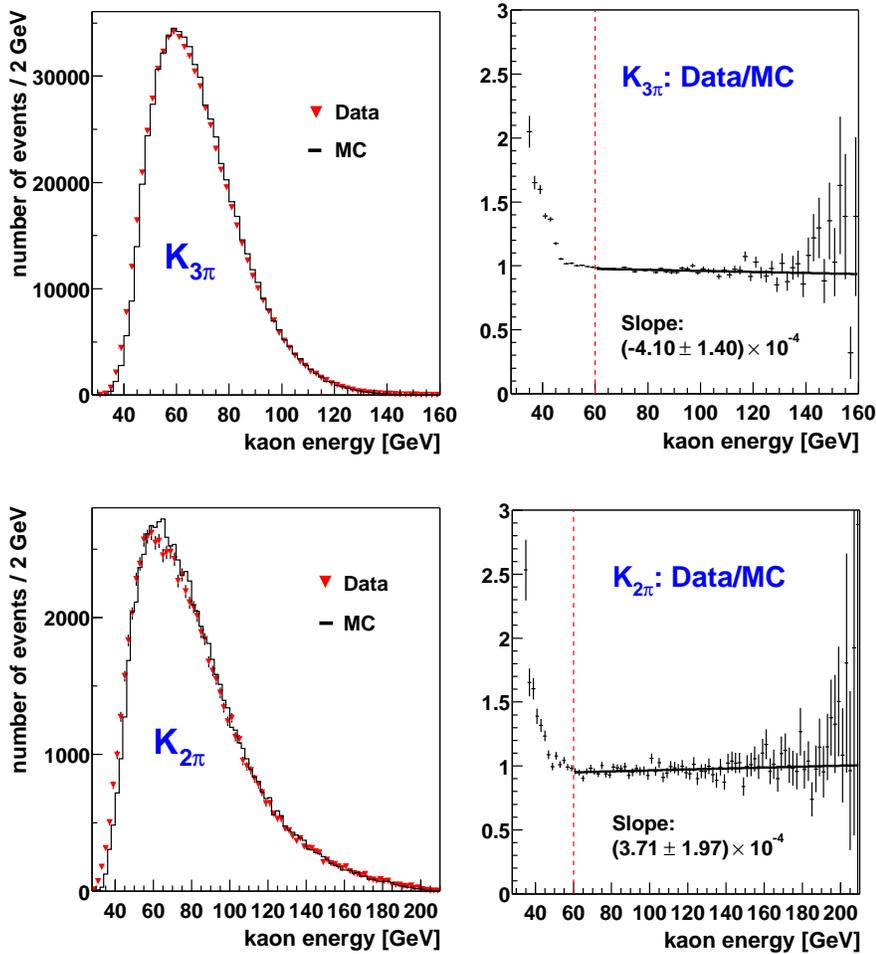}
    \parbox{12cm}{\caption[]
      {\labf{ekaon_k3pi_k2pi}Reconstructed kaon energy from $K_{3\pi}$
      and $K_{2\pi}$ decays, comparison between data and MC. Errors are statistical only.}}
  \end{center}
\end{figure}
In \Fig{ekaon_k3pi_k2pi} we show a comparison of the energy spectra for identified
$K_{3\pi}$ and $K_{2\pi}$ events, where we can fully reconstruct the
energy. Both comparisons show a small slope but with opposite signs,
demonstrating that the kaon energy spectrum in the MC is a good
compromise between different decay modes.

\Fig{ke3_psum_el_pi} compares the momenta
of the electrons and pions in
identified $K_{e3}$ events between data and MC. \Fig{ke3_psum_zoom}
shows the same comparison for the sum of track momenta in the range
between 60\,GeV/c and 130\,GeV/c, which contains 95\,$\%$ of the
data. With radiative corrections applied, we still observe a slight
slope in $P$ of half the size of the slopes of the fully reconstructed
events in \Fig{ekaon_k3pi_k2pi}. This is the dominant source of
experimental uncertainty, and may be due to imperfections of the
radiative corrections as well as limitations in the detailed event simulation.
\begin{figure}[h]
  \begin{center}
    \includegraphics[width=12.5cm]{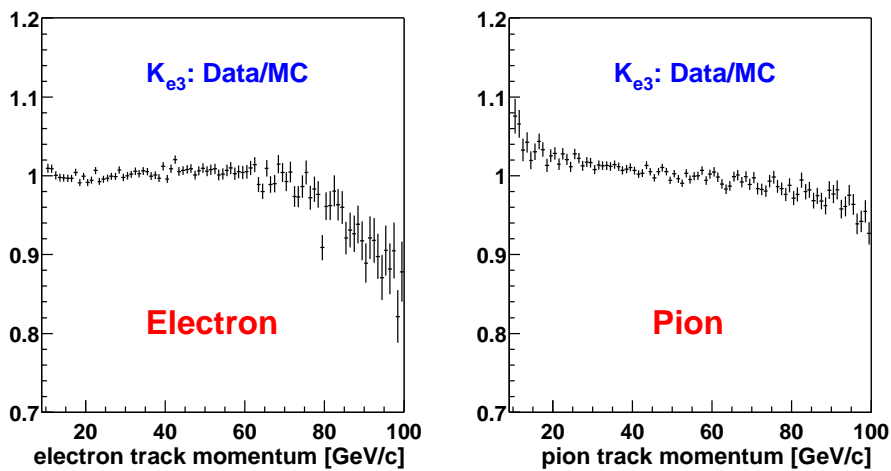}
    \parbox{12cm}{\caption[]
      {\labf{ke3_psum_el_pi}Comparison between data and MC (including
      radiative corrections) for the momenta of electrons and pions in
      identified $K_{e3}$ events. Errors are statistical only.}}
  \end{center}
\end{figure}
\begin{figure}[h]
  \begin{center}
    \includegraphics[width=10cm]{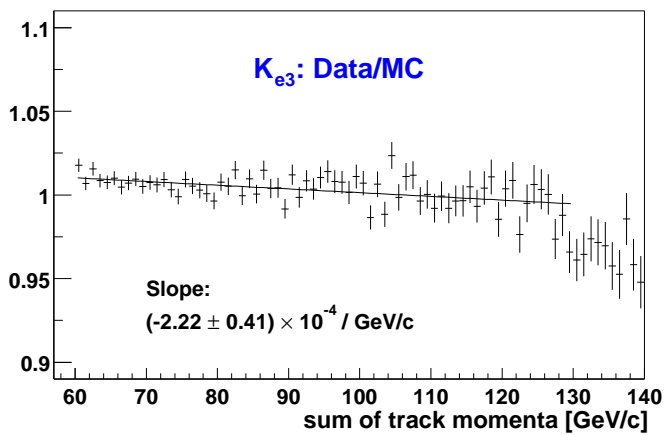}
    \parbox{11cm}{\caption[]
      {\labf{ke3_psum_zoom}Comparison between data and MC for the sum
      of track momenta. The region between 60\,GeV/c and 130\,GeV/c
      contains 95\,$\%$ of the data. Errors are statistical only.}}
  \end{center}
\end{figure}

To get a conservative estimate of this dependence, we varied the lower
cut on the value of $P$ from 50\,GeV/c to 80\,GeV/c. This is a large 
range of variation, considering that the analysis used
data above 60\,GeV/c, and that a cut at 80\,GeV/c removes
70\,$\%$ of the events. The resulting relative uncertainty of the ratio
$R$ in Eq. (\ref{eq:R}) is 0.67\,$\%$. In addition, a second independent analysis
was performed using different selection criteria and a different kaon
momentum spectrum, which was weighted such as to reproduce exactly the
kaon momentum spectrum of $K_{e3}$ events. The value of $R$ differed from the
one in the first analysis by 0.2\,$\%$, well below the estimated
systematic uncertainty.

To estimate the uncertainty coming from the $E/p$ cut to select $K_{e3}$
events, we varied the cut value between $E/p$ $> 0.90$ and $E/p$ $> 0.96$. As
a result, inefficiency and background due to this criteria vary
significantly, leading to different net corrections of $K_{e3}$
event numbers (Table \ref{tab:eopCutVariations}). Applying these corrections, 
however, we get almost the same number of events, thus demonstrating the 
correctness of this selection principle. It appears that with $E/p$ $>
0.93$, both inefficiency and background are very small and nearly cancel.
The resulting relative uncertainty on $R$ is $\Delta R / R = 0.05\,\%$.
\begin{table}[ht]
  \begin{center}
    \begin{tabular}[t]{|l|c|c|c|}\hline
      & $E/p$ $> 0.90$ & $E/p$ $> 0.93$ & $E/p$ $> 0.96$\\	\hline
      inefficiency [$\%$] & 0.275 & 0.487 & 1.424\\ 
      background [$\%$] & 0.914 & 0.576 & 0.266\\ 
      $K_{e3}$ event number after $E/p$ cut& 6796461 & 6759184 & 6673114\\
      net $K_{e3}$ correction & -42624 & -5705 & 77182\\
      corrected $K_{e3}$ event number & 6753836 & 6753478 & 6750296\\ \hline
    \end{tabular}
  \end{center}
  
  \caption{Variation of the $E/p$ cut to select $K_{e3}$ events.}
  \label{tab:eopCutVariations}
\end{table}

The data used in this analysis originate from two different
triggers $(Q2 + Q1/20*2trk)$, where $Q2$ requires two quadrants of 
the hodoscope counter to be hit, while $Q1*2trk$ requires at least one 
hodoscope quadrant plus two tracks from the drift chamber trigger
system. Q1 is prescaled by a factor of 20. By selecting one trigger, the efficiency of the other
can be measured, taking into account the different downscaling. 
The trigger efficiencies for 2-track and $K_{e3}$ events differ
slightly for the Q2 trigger ( $(97.38 \pm 0.02)\,\%$ for 2-track
events, $(97.49 \pm 0.03)\,\%$ for $K_{e3}$ events). As a check, the
analysis was repeated for the $Q1*2trk$ trigger alone,
which was measured to be equally efficient for all events. 
The relative uncertainty due to different trigger efficiencies 
is very small: $\Delta R / R = 0.05\,\%$. 

In about 5\,\% of the events the drift chambers record multiple hits in one layer which
lead to an overflow condition. This could be more likely
for electrons than for minimal ionizing pions or muons. Comparing the
results with or without cutting on the overflow condition shows that
the effect on $R$ is almost negligible: $\Delta R / R = 0.05\,\%$.  

Using a data set of monochromatic single pions or electrons from a
test run, it has been checked that the efficiencies to record and
reconstruct pions and electrons are equal within 0.05\,\%.

In order to be independent of potential asymmetries in the setup, about half of the 
data were recorded with positive polarity and half with negative
polarity of the spectrometer magnet. We analyzed the data separately
for both polarities, but found an almost negligible difference, resulting in 
an uncertainty of $\Delta R / R = 0.07\,\%$.

As a further systematic check the analysis was repeated, broadening a
number of detector resolutions in Monte Carlo. Energy, momentum and
vertex positions were convoluted with gaussian distributions, the
chosen standard deviations being half of the experimental resolution.
The number of selected events changed only by
the order of $10^{-5}$, proving that the result does not depend on
resolution effects. 
\begin{table}[ht]
    \begin{center}
      \begin{tabular}[t]{|l|c|} \hline
	& relative uncertainty [$\%$]\\ \hline
	experimental normalization (energy spectrum) & 0.67 \\ 
	normalization error from input ratios & 0.16\\ 
	$E/p$ cut & 0.05\\ 
	trigger efficiency & 0.05\\
	DCH overflows & 0.05\\  
	magnet polarity & 0.07\\ \hline
      \end{tabular}
    \end{center}    
    \caption{Summary of systematic uncertainties on the ratio $R$.}
    \label{tab:systUncertainties}
\end{table}

We summarize the systematic uncertainties on $R$ in Table \ref{tab:systUncertainties}. 
Using the acceptances given in Table 1, the ratios of branching
fractions from Eq. (\ref{eq:ExR1}) to (\ref{eq:ExR11}) and the above
evaluation of the systematic uncertainties, we obtain as average two
track acceptance $a_{2T} = 0.2412 \pm 0.0004$, and a systematic uncertainty in the ratio
$a_e/a_{2T}$ of $0.68\,\%$.

%
\section{Results}

The electron identification inefficiency increases the number of
$K_{e3}$ decays by $0.49 \%$, while background from
misidentified $K_{\mu3}$ and $K_{3\pi}$ decays reduces the number by
$0.58 \%$, leading to a net correction of -5705 events. This gives:
\begin{eqnarray}
R = \frac{B(K_L \rightarrow \pi e \nu)}{B(K_L \rightarrow
  all~\mbox{2-}track) )}= \frac{6753478/0.2599}{12592096/0.2412} = 0.4978 \pm
0.0035\,. \label{eq:resultR}
\end{eqnarray}

As mentioned previously, a second independent analysis resulted in a value
of $R$ which differs by less than 0.001 from the value in Eq. (\ref{eq:resultR})\,.

For the branching ratio of the $3 \pi^0$ decay, the current
experimental situation is unsatisfactory. We use a weighted mean of
the PDG2004 value $(21.05 \pm 0.28)\,\%$ \cite{PDG} and the recent
measurement of the KTeV collaboration, $(19.45 \pm 0.18)\,\%$ \cite{KTEV},
and obtain $(19.92 \pm 0.70)\,\%$, where the error is enlarged because
of the poor agreement of the measurements. Therefore the branching
ratio for all 2-track events is $B(2T)=(80.56 \pm 0.70)\,\%$ and 

\begin{eqnarray}
B(e3) = \frac{\Gamma(K_L \rightarrow \pi e \nu)}{\Gamma(K_L
\rightarrow all)}= R \ast B(2T) = 0.4010 \pm 0.0028 \pm 0.0035 ,
\end{eqnarray}

with the first error being the complete experimental error and the
second the external error from the normalization, to be combined to
\begin{eqnarray}
B(e3)= 0.4010 \pm 0.0045 .
\end{eqnarray}

This measurement depends on three other measurements of ratios of
partial $K_{e3}$ decay widths. This dependence is given by:

\begin{eqnarray}
\Delta B(e3)= \left(\frac{\Gamma(\mu3)}{\Gamma(e3)}-0.666\right)*0.077 & - &
\left(\frac{\Gamma(3\pi)}{\Gamma(e3)}-0.309\right)*0.075 \nonumber \\
 & - & \left(\frac{\Gamma(3\pi^0)}{\Gamma(e3)}-0.515\right)*0.151.
\end{eqnarray}

The decay rate of $K_L \rightarrow \pi e \nu$ is obtained by using the
$K_L$ lifetime $\tau(K_L) = (5.15 \pm 0.04)\times 10^{-8}\,\mbox{s}$ \cite{PDG}:
\begin{eqnarray}
\Gamma(K_{e3}) = B(e3)/\tau(K_L) = (7.79 \pm 0.11)\times 10^6\,\mbox{s}^{-1} .
\label{Gammae3}
\end{eqnarray}

%
\section{Value of $|V_{us}|$}

The CKM matrix element $|V_{us}|$ can be extracted from the $K^0_{e3}$
decay parameters by ref. \cite{Ciri}
\begin{eqnarray}
|V_{us}|=
 \sqrt{\frac{128\pi^3\Gamma(K^0_{e3})}{G^2_FM^5_{K^0}S_{EW}I_{K^0}}}\frac{1}{f^{K^0
 \pi^-}_{+}}
\end{eqnarray}

Three quantities in this equation are taken from
theory. $S_{EW}$ is the short distance enhancement factor, $I_{K^0}$
is the phase space integral and $f^{K^0 \pi^-}_{+}$ is the form factor.

To determine $|V_{us}|$ we follow the prescription and use the numerical
results in ref. \cite{Ciri}, where a detailed numerical
study of the $K_{e3}$ decays to $\mathcal{O}(p^6)$ in chiral
pertubation theory with virtual photons and leptons is presented. The integrals
given therein correspond to the specific prescription to accept only those 
radiative events which have pion and electron
energies within the whole $K_{e3}$ Dalitz plot. From a Monte Carlo
simulation we obtain this correction to be small
\begin{eqnarray}
\frac{\mbox{Number of $K_{e3(\gamma)}$ events inside Dalitz
plot}}{\mbox{Number of all $K_{e3(\gamma)}$ events}} = 0.99423\,.
\label{Rad}
\end{eqnarray}

Using equations (\ref{Gammae3}) and (\ref{Rad}), $S_{EW}=1.0232$,
$I_{K^0}=0.10339\pm0.00063$ we obtain a value for the product of the CKM matrix
element $|V_{us}|$ and the vector form factor $f^{K^0 \pi^-}_{+}$,

\begin{eqnarray}
|V_{us}|f_{+}(0) = 0.2146 \pm 0.0016 \,.
\end{eqnarray}

For the vector form factor, different theoretical calculations have been published recently. 
Chiral models including the corrections to the order $p^6$ give
$f_{+}(0) = 0.981 \pm 0.010$ \cite{Ciri},
$f_{+}(0) = 0.976 \pm 0.010$ \cite{Bijnens} and $f_{+}(0) = 0.974 \pm
0.011$ \cite{Jamin}, to be compared with the older value $f_{+}(0) =
0.961 \pm 0.010$ \cite{Roos}. Lattice calculations in the quenched
fermion approximation give $f_{+}(0) =
0.961 \pm 0.009$ \cite{Beci}, but this value does not include
electromagnetic corrections. Taking the value from
\cite{Ciri}, which takes into account chiral corrections to the order
$p^6$, isospin corrections and electromagnetic corrections, we obtain
the CKM element to be

\begin{eqnarray}
|V_{us}| = 0.2187 \pm 0.0028 \,.
\end{eqnarray}

The error on $|V_{us}|$ is dominated by the theoretical uncertainties,
the error on $f^{K^0 \pi^-}_{+}$ alone contributing $\pm0.0023$.

%
\section{Conclusions}

We have made a direct measurement of the ratio of $K^0_{e3}$ to all
$K^0_L$ decays with two charged tracks,
\begin{eqnarray}
R = \frac{B(K_L \rightarrow \pi e \nu)}{B(K_L \rightarrow
  all~\mbox{2-}track)} = 0.4978 \pm 0.0035\,.
\end{eqnarray}

Using the current experimental knowledge of the $3 \pi^0$ branching ratio,
this leads to a branching ratio $B(e3)= 0.4010 \pm 0.0045$. This exceeds
the PDG value by ($3.3 \pm 1.3$)\,\%, or 2.5 standard deviations. It leads
to $|V_{us}|f_{+}(0) = 0.2146 \pm 0.0016$, in good agreement with the
recent KTeV result \cite{KTEV}, but larger than the PDG value
\cite{PDG}. Inferring the most recent theoretical evaluation of
$f_{+}(0) = 0.981 \pm 0.010$ \cite{Ciri}, the coupling
constant comes out to be $|V_{us}| = 0.2187 \pm 0.0028$, where the dominant
uncertainty is theoretical. This is still 2.4 sigma lower than
required by the 3-generation unitarity of the CKM matrix.

\section{Acknowledgements}

We gratefully acknowledge the continuing support of the technical
staff of the participating institutes and their computing centers.

\end{document}